# Chiral light-matter interactions using spin-valley states in transition metal dichalcogenides

**Zhili Yang[1*], Shahriar Aghaeimeibodi[1] and Edo Waks[1,2*]**

[1] *Institute for Research in Electronics and Applied Physics, University of Maryland, College Park, Maryland 20742, USA*

[2] *Joint Quantum Institute, University of Maryland and National Institute of Standards and Technology, College Park, Maryland 20742, USA*

[*yangzl@umd.edu](mailto:yangzl@umd.edu), [*edowaks@und.edu](mailto:edowaks@und.edu)

**Abstract:**

Chiral light-matter interactions can enable polarization to control the direction of light emission in a photonic device. Most realizations of chiral light-matter interactions require external magnetic fields to break time-reversal symmetry of the emitter. One way to eliminate this requirement is to utilize strong spin-orbit coupling present in transition metal dichalcogenides that exhibit a valley dependent polarized emission. Such interactions were previously reported using plasmonic waveguides, but these structures exhibit short propagation lengths due to loss. Chiral dielectric structures exhibit much lower loss levels and could therefore solve this problem. We demonstrate chiral light-matter interactions using spin-valley states of transition metal dichalcogenide monolayers coupled to a dielectric waveguide. We use a photonic crystal glide plane waveguide that exhibits chiral modes with high field intensity, coupled to monolayer $WSe_2$. We show that the circularly polarized emission of the monolayer preferentially couples to one direction of the waveguide, with a directionality as high as 0.35, limited by the polarization purity of the bare monolayer emission. This system enables on-chip directional control of light and could provide new ways to control spin and valley degrees of freedom in a scalable photonic platform.

Chiral photonic structures couple the spin of photons from an emitter to the direction of propagation[1–3]. This coupling leads to chiral light-matter interactions where the polarization of the emitter controls the flow of light, which could enable directional spin-photon interfaces[4,5] and polarization-controlled optical transistors and circulators[2,6]. Such chiral light-matter interactions have been previously realized using atoms coupled to fibers and resonators[7–9], as well as quantum dots[10,11]. However, these previous realizations require large magnetic fields to break time-reversal symmetry. But the requirement for large magnetic fields makes it complicated to integrate these components onto a compact chip.

Transition metal dichalcogenides open up the possibility to attain chiral light-matter interactions without magnetic fields[12]. These materials exhibit valley-dependent optical selection rules, where the excitation source determines the polarization of the emission[13–15]. When coupled to chiral photonic structures, the emission can exhibit chiral light-matter interactions controlled by the input polarization, without the need for magnetic fields. Recent work has demonstrated this type of chiral light-matter interaction using $WS_2$ coupled to plasmonic waveguide[16,17]. Dielectric chiral waveguides provide an alternative approach[18]. These waveguides exhibit significantly lower losses than plasmonic structures, and can simultaneously attain strong chirality and field intensity at the same location[4]. Dielectric photonic crystal waveguides can have losses as low as dB per centimeter[19] providing a propagation length much longer than what is possible with plasmonic waveguides at optical frequencies [20]. However, such chiral interactions between transition metal dichalcogenide monolayers and dielectric photonic structures has not been realized.

Here we demonstrate chiral light-matter interactions using transition metal dichalcogenide monolayers coupled to a chiral dielectric waveguide. We use $WSe_2$ monolayers coupled to a glide-plane photonic crystal waveguide, which supports strong chiral light-matter interactions due to broken mirror symmetry[4]. We demonstrate directional control of light transmission of the $WSe_2$ emission by controlling the input polarization without an external magnetic field. This chiral interface could open up new approaches for information processing by using the spin-valley degree of freedom in strong spin-orbit materials coupled to dielectric integrated photonic circuits[21].

Figure 1a shows a scanning electron microscope image of the glide plane waveguide. This structure, originally proposed by Kuang and O'Brien[22], is composed of a photonic crystal waveguide where two top and bottom mirrors are shifted by a half period. This shift breaks mirror symmetry, resulting in optical modes with in-plane circular polarization that enable strong light-matter coupling[4]. Previous work on chiral light-matter coupling in glide-plane waveguides utilized GaAs as the photonic substrate. But GaAs is not transparent at the emission wavelength of transition metal dichalcogenides. We therefore adapt the glide-plane waveguide design to SiN, which is highly transparent over the visible and near infrared wavelengths. We consider a SiN membrane with a refractive index of $n = 2.01$ and a thickness of $t = 200$ nm. We set the photonic lattice constant a = 290 nm and the hole radius $r = 80$ nm.

Figure 1b shows simulated electric field intensity profiles for the designed glide-plane photonic waveguide in response to a dipole emitter with right and left circular polarization located at the field maximum. We performed these calculations using finite-difference time-domain simulation. In the calculations, we used a dipole point excitation source with a frequency contained within the propagation band of the waveguide. The light coupled to the waveguide exhibits unidirectional propagation, where the direction depends on the polarization of the dipole. This unidirectional propagation is the signature behavior of chiral light-matter coupling. We also simulated the coupling efficiency of the emission from the dipole to the waveguide mode for both polarizations to be around 30%. (See supplementary figure 1)

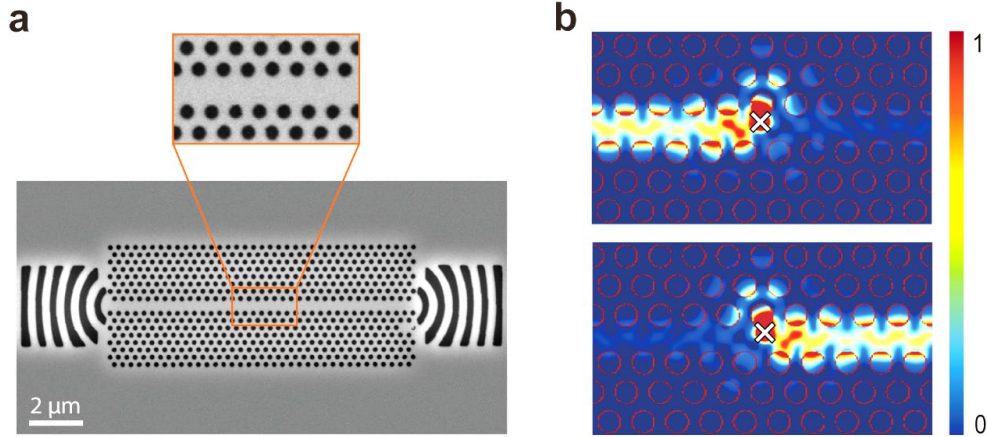

**Figure 1.** (a) Scanning electron micrograph of the fabricated chiral photonic waveguide with grating couplers at both ends. (b) Simulated electric field intensity profiles in the glide-plane photonic crystal waveguide excited by right (top) and left (bottom) circularly polarized dipoles. We mark the position the dipole by the white cross.

To experimentally demonstrate chiral light-matter coupling we need to add a circularly polarized dipole emitter to the fabricated device, since the photonic structure itself is time-reversal symmetric. One way to realize such an emitter is to exploit the Zeeman-split optical transitions of a quantum emitter such as a quantum dot[4,10,11]. But this approach requires a high magnetic field. Transition metal dichalcogenides provide an alternative approach. These materials exhibit circularly polarized emission when excited by a circularly polarized pump[13,23], even without a magnetic field. Thus, we can directly control the emission polarization by using pump light with an appropriately prepared polarization state. We note that because the glide-plane waveguide modes are below the light-line, it is not possible to excite the waveguide directly with the laser from the out-of-plane direction. Such a process would not conserve momentum.

In this work we use monolayer $WSe_2$ which emits at a wavelength of 710-750 nm. We first grow the monolayers using standard chemical vapor deposition[24]. We then transfer the monolayers using a Polydimethylsiloxane (PDMS) gel as a transfer stamp[25]. Figure 2a shows a camera image of the fabricated device, where the triangular dashed line indicates the outline of the monolayer, which overlaps with the glide plane waveguide.

To perform measurements, we cool down the sample to 5 K using a closed-cycle refrigerator. We perform all measurements using a confocal microscope system with an objective lens that has a numerical aperture of 0.7. We excite the sample either with a 633 nm laser diode, or with a Ti:sapphire laser tuned to 710 nm, which is close to the exciton emission wavelength of $WSe_2$. We control the polarization of excitation and collection ports independently using different quarter and half waveplates. A 725 nm long-pass filter rejects the direct laser reflection at the output port, isolating only the fluorescence signal from the monolayer.

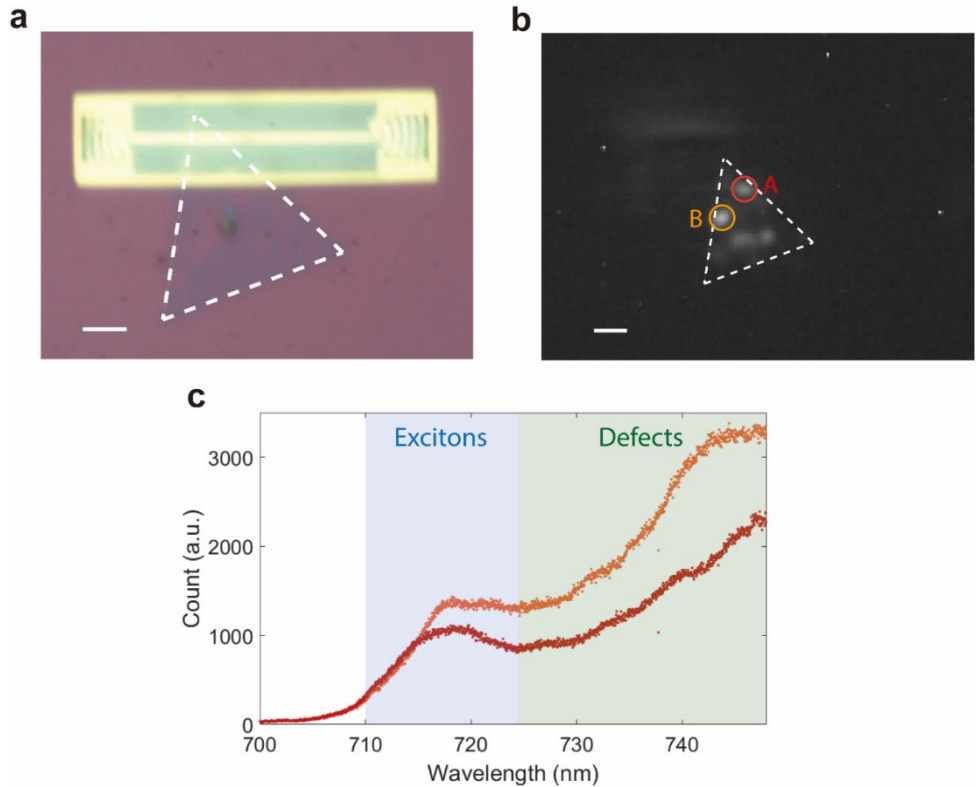

**Figure 2.** (a) Optical image and (b) photoluminescence map of a fabricated device. Dotted triangle indicates the area of a large WSe$_2$ monolayer flake. The red (yellow) circle indicates the local emission point A (B) for the measurements. Scale bar is 2 μm. (c) photoluminescence spectrum from point A (red) and point B (yellow) in (b).

Figure 2b shows the photoluminescence map where we excite the sample using the 633 nm laser. The photoluminescence intensity is highly non-uniform and localized at several diffraction limited spots on the membrane. This localized emission is consistent among virtually all fabricated devices, and is typical in monolayer WSe$_2$ as reported by a number of other works[26–28]. We identify two points of interest in the image, point A which corresponds to localized emission on the waveguide and point B which correspond to a localized emission on the unpatterned SiN substrate. Figure 2c shows the photoluminescence spectrum from these two points. In both cases the emission exhibits a broad spectrum with two peaks, one corresponding to the exciton line and the other to the defect-bound exciton emission.[29–32] We note that the spectrum does not show sharp bright emission lines consistent with single photon emitters[33,34]. We therefore do not attribute the localized emission spots in figure 2b to single defects.

To explore the chirality of the light emission, we excite the coupled monolayer flake with a circularly polarized 710 nm excitation laser that pumps the exciton line of WSe2, and collect the output defect-bound photoluminescence (725 nm to 750 nm) from the right and then left gratings as illustrated in Figure 3a. Figure 3b and 3c show the spectra collected from the right and left grating respectively for both right and left circularly polarized pumps. In the right grating, we observe a stronger signal for right circularly polarized light as opposed to left circular polarization. The left grating shows precisely the reverse behavior, where we obtain higher signal using a left circularly polarized pump. This asymmetry is a strong evidence of chiral light-matter coupling. We note that the spectrum features periodic ripples due to Fabry-Perot oscillations caused by partial reflection from the two ends of the waveguide. Also, both spectra exhibit a cutoff at wavelengths below 725 nm due to the long pass filter we use to reject the pump. Similar spectra for another device under 633 nm excitation are provided in Supplementary Figure 2.

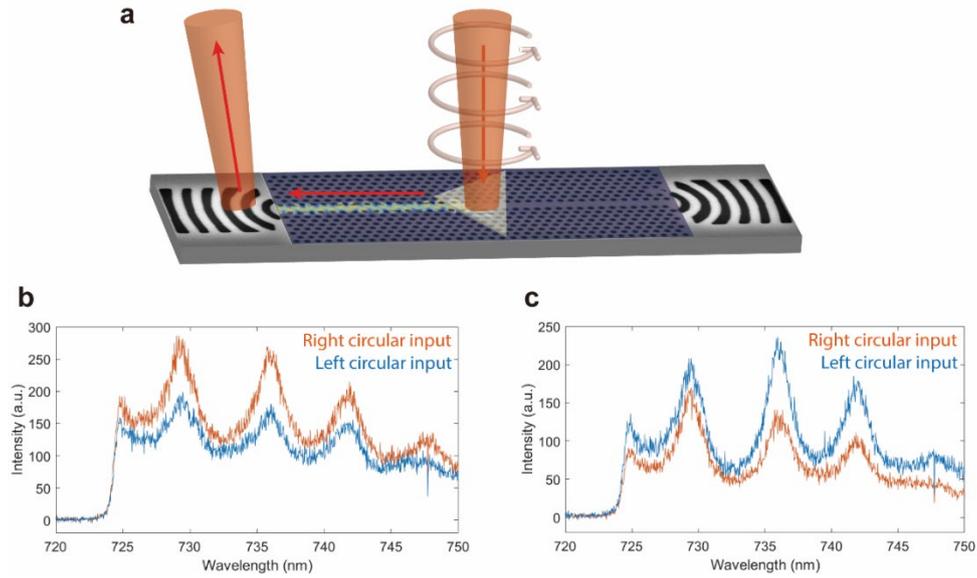

**Figure 3.** (a) Schematic of the chiral light-matter interaction. We excited the coupled monolayer with circularly polarized excitation and collected the signal at the right and then the left grating end. Emission spectra collected from (b) right and (c) left ends of the waveguide under orthogonal circularly polarized pump excitation. Orange (blue) curves indicate emission under right (left) circular polarization.

We define the directionality of the emission as

$$F(\lambda) = \frac{I_R(\lambda) - I_L(\lambda)}{I_R(\lambda) + I_L(\lambda)}$$

where $I_R(\lambda)$ and $I_L(\lambda)$ are the output intensity under right and left circularly polarized excitation at wavelength $\lambda$, respectively. Figure 4a shows the calculated directionality for the data in Figure 3b-c. To reduce the noise, we average over 0.5 nm wavelength bins, which is significantly narrower than all spectral features in the measurement. We observe consistent positive (negative) directionality over the entire WSe$_2$ emission spectral range at the right (left) end of the device. The sign change of the directionalities between the right and the left ends indicates that the circular polarization of the input can modulate the direction of the emission from the monolayer. The directionality depends on a combination of the degree of polarization of the emitter and it is location relative to the chiral waveguide. At the optimal chiral point we expect the directionality to be determined by the degree of polarization alone which lies between 0.3 to 0.4 (see Supplementary Figures 3-4 and Ref. [23]). But the actual chirality will be lower if the emitter is not located precisely at this optimal point. In all cases, our measurements lie within the allowable range set by the degree of polarization.

The directionality of the coupled device depends on the position of the localized emission relative to the mode of the glide-plane waveguide. Figure 4b shows the calculated directionality for a left circularly polarized emitter as a function of position on the glide plane waveguide. The directionality can be both positive or negative depending on the location of the emitter. A linearly polarized emitter on the other hand will not exhibit directionality (see Supplementary Fig. 5). Since the emitters in our device are randomly dispersed on the surface, we expect that emitters will exhibit different directionality. To validate this assertion, we measure 21 localized emitters in 18 coupled devices. We define $F_{max}$ as the maximum directionality over the emission wavelength range. Figure 4c plots a histogram of $F_{max}$ over all measured localized emission points. The directionality shows a nearly uniform distribution over the range of -0.4 to 0.4, as expected from the distribution of the directionalities in Fig. 4b and the degree of polarization of the monolayer. We note that the position dependence of the chiral interaction is not unique to our structure. It is a feature of all chiral coupled systems, which require a strong field gradient. The glide plane waveguide has the advantage that it aligns the chiral point and the field maximum in the same place, allowing strong chiral light-matter coupling. Pre-patterning the substrate could provide a more systematic way to align the emission of the monolayer with the chiral modes of the waveguide.

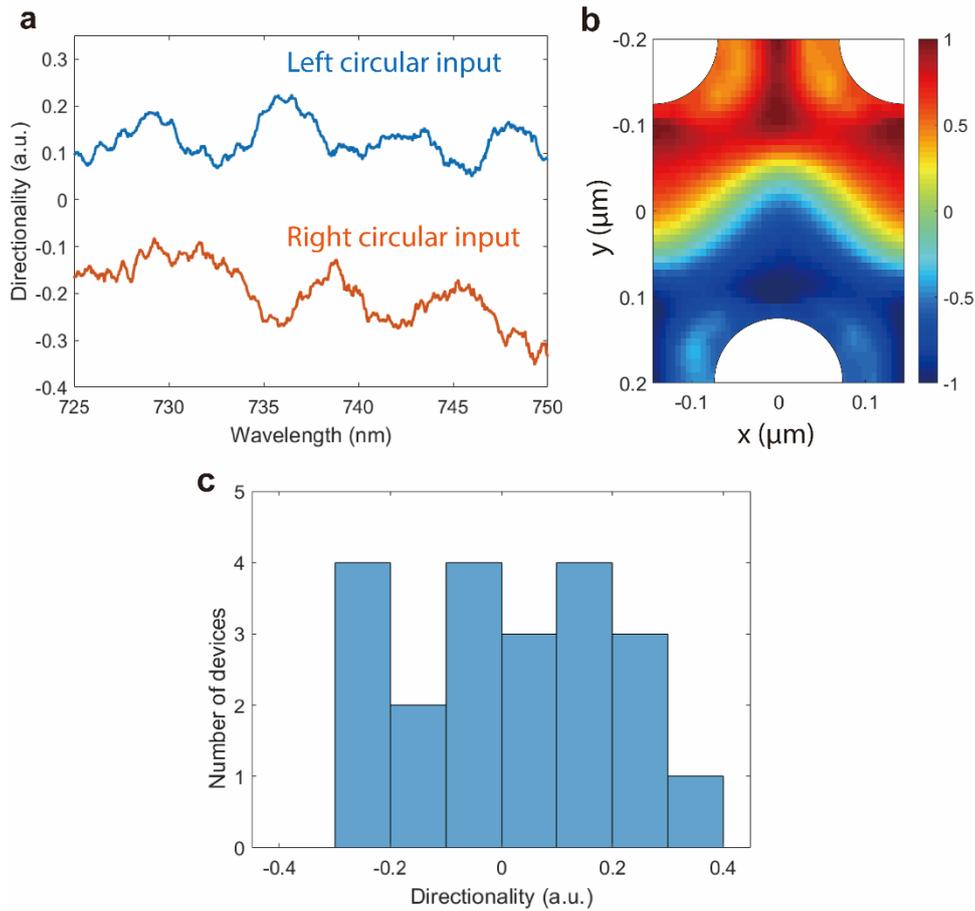

**Figure 4.** (a) Directionality as a function of wavelength. (b) Calculated directionality for a left circularly polarized emitter as a function of position on the glide plane waveguide. (c) Distribution of directionality for 21 measured devices.

In conclusion, we demonstrate chiral light-matter interactions by coupling $WSe_2$ monolayers to a chiral photonic crystal waveguide structure. Using this structure, we can control the emission direction using the polarization of the excitation pump without the need for strong magnetic fields. The directionality is consistent with the expected values based on the degree of polarization of the bare monolayer. Even higher directionality and room temperature operation could therefore be possible using other atomically thin materials that support higher degree of polarization[13]. Employing more complex asymmetric waveguides such as gammadion structures[35] may provide a higher directionality. Active alignment of emitters with the waveguide could enable deterministic alignment of the emission with the chiral mode[36]. In addition, these materials can also contain single photon emitters that can be coupled to glide-plane waveguides[37,38], providing a directional single photon source. Ultimately, atomically thin materials coupled to glide-plane waveguides could serve as building blocks for integrated photonic devices that can control and route light using polarization[39–41].

**Funding:**

The authors acknowledge support from an Office of Naval Research (ONR) (N00014172720), the Air Force Office of Scientific Research (AFOSR) (FA9550-18-1-0161), and the National Science Foundation (NSF) (ECCS1508897).